\documentclass{svproc}
\usepackage{url}

\usepackage{color}
\usepackage{soul}
\usepackage{algorithm}
\usepackage[noend]{algpseudocode}
\usepackage{amsmath}
\usepackage{graphicx}
\usepackage{amssymb}

\algdef{SxnE}[REPEATN]{RepeatN}{End}[1]{\algorithmicrepeat\ #1 \textbf{times}}{}

\begin{document}
\mainmatter              % start of a contribution
\title{A simple and flexible algorithm to generate\\ real-world networks}
\titlerunning{A simple and flexible algorithm to generate real-world networks}  % abbreviated title (for running head) also used for the TOC unless \toctitle is used
\author{João Pedro C. Morais \and Ruben Interian} % \inst{1}
\authorrunning{J. P. C. Morais and R. Interian} % abbreviated author list (for running head)
%
%%%% list of authors for the TOC (use if author list has to be modified)
\tocauthor{João Pedro C. Morais and Ruben Interian}
\institute{Instituto de Computação,\\ 
Universidade Estadual de Campinas (UNICAMP)\\ 
Campinas 13083-852, São Paulo, Brazil\\
\email{ruben@ic.unicamp.br}}

\maketitle % typeset the title of the contribution

\begin{abstract}
This study introduces an algorithm that generates undirected graphs with three main characteristics of real-world networks: scale-freeness, short distances between nodes (small-world phenomenon), and large clustering coefficients. 
The main idea is to perform random walks across the network and, at each iteration, add special edges with a decreasing probability to link more distant nodes, following a specific probability distribution.
%This results in a simple algorithm with low time complexity, generating networks with desirable features. 
A key advantage of our algorithm is its simplicity and flexibility in creating networks with different characteristics without using global information about network topology. 
We show how the parameters can be adjusted to generate networks with specific average distances and clustering coefficients, maintaining a long-tailed degree distribution. The implementation of our algorithm is publicly available on a GitHub repository. 

% We encourage you to list your keywords within the abstract section using the \keywords{...} command. 
\end{abstract}

\section{Introduction}

% https://complenet.weebly.com/call-for-contributions.html

In recent years, the study of complex networks has gained prominence across various scientific disciplines, such as sociology, physics, biology, and computer science~\cite{2018DeArruda,2023Interian,2023Seguin,2023Zhou,2024Interian}. Real-world networks often display both scale-free characteristics and surprising proximity among network nodes. These two features commonly observed in real-world networks are the so-called Matthew effect and the small-world phenomenon. 

The Matthew effect (also known as ``rich-get-richer effect'', or accumulated advantage) reflects a preferential attachment dynamic~\cite{1999Barabasi}, leading to the emergence of high-degree hubs in a network and long-tailed degree distributions~\cite{2019Broido}. 
%according to $p(k) \propto k^{-\gamma}$, where $p(k)$ is the proportion of nodes with degree k. 
This behavior can be easily observed in social networks, where most people have few connections, and few people hold a very large number of connections, showing a long-tail pattern.~\cite{BHATTACHARYA20201200} 

Social networks are also examples of networks with high clustering, where individuals tend to form tightly connected groups, such as friend circles, family groups, or professional communities. In these clusters, the likelihood of any two friends of some individual also being friends is higher than in random networks, resulting in a high clustering coefficient.   

Finally, the small-world property characterizes networks with short average path lengths, enhancing navigability. 
Collaboration networks of actors and the neural network of the nematode worm \emph{C. elegans} are examples of small-world networks~\cite{1998Watts}. 

Modeling networks with these three characteristics can be challenging, as the mechanisms underlying each of them tend to drive network structures in different directions. 
Random walk models~\cite{Saram_ki_2004} were previously used to generate networks with long-tailed degree distributions, capturing real-world systems' natural growth and preferential attachment. 
A key advantage is their ability to create structures without global information, aligning with the organic development of real networks. 
While effective at generating hubs and modeling the Matthew effect, traditional random walks often overlook reducing path lengths between nodes -- a crucial feature for small-world networks. 
This limitation makes it challenging to model systems requiring long-tailed degree distributions, high clustering, and short average distances. 

In this paper, we present a random walk algorithm designed to generate networks that show small-world features, Matthew effect, and high local clustering, 
%employing clustering control techniques on the walks, while 
adding special edges that emulate randomness in link generation observed in real networks, thus reducing the average distances between the nodes in the network. 

The paper is organized as follows. Section 2 presents previous works that inspired and contributed ideas that led to the presented approach. In Section 3, we describe our algorithm. The results and characteristics of the generated networks are presented in Section 4. Conclusions are detailed in the last section. 

\section{Previous works}

Previous works have studied the generation of networks with specific combinations of features considered in our study. 

Arguing that real-world networks are often highly clustered while showing small average distances between nodes, Watts and Strogatz~\cite{1998Watts} proposed a model to reproduce such characteristics. Starting with a set of $N$ nodes in a circular order where each node is connected with an undirected edge to $k$ neighbors, the authors rewired each edge with some fixed and small probability $p$. The average clustering coefficient remained quite high while the average distance dropped to a small value approximately proportional to $\log N$. 

Latent-space network models have also been employed to investigate small-world networks with non-vanishing clustering~\cite{Bogu2021}. 
In its simplest version, also called random geometric graph model, nodes are distributed uniformly at random in some metric space, and two nodes $i$ and $j$ are connected if and only if the distance $x_{i j}$ between them is less than some parameter $\mu$, leading to high clustering and large average shortest paths. 
% that scales linearly with network size (these networks are large worlds). 
However, the model can become small-world by introducing a probability $p_{i j}$ of the existence of a link between the nodes. 
For example, in $\mathbb{R}^d$ space, choosing $p_{i j} \propto x_{i j}^{-\beta}$ with $\beta \in (d, 2d)$ results in non-vanishing clustering coefficients and small-world networks, with average distances scaling proportionally to $\log N$~\cite{Bogu2020}. 

On the other hand, real-world networks also have high-degree nodes called hubs, which are absent in the above models. 
%Watts and Strogatz' model and in latent-space network models. 
Barabási and Albert~\cite{1999Barabasi} proposed a preferential attachment process that generates such long-tailed degree distributions. 
Starting from some small graph, one new node $v$ is added at each iteration with $k$ new edges linking $v$ to $k$ different nodes chosen with probabilities proportional to node degrees. That is, the likelihood of node $v$ choosing node $w$ is proportional to the degree of $w$ at that iteration, generating scale-free networks.

To increase the clustering coefficient of the networks generated by the above BA model, Holme and Kim~\cite{2002Holme} introduced a Triad Formation step performed with probability $P_t$ after a node and its edges are added to the network. When a new edge is added linking the new node $v$ to a node $w$, an edge is added linking $v$ to a randomly selected neighbor of $w$, thus creating a triad between the three nodes and increasing the clustering of the network. 

Although previous models have presented solid results in scale-free network construction, a significant concern we raise is that they require global information at each step (e.g., the degrees of all nodes to calculate the preferential attachment probabilities), which may be unrealistic since in real-world networks links emerge naturally, without knowing global information about network topology. 

Saramäki and Kaski~\cite{Saram_ki_2004} proposed using random walkers to generate undirected scale-free networks, showing that it is not necessary to have global information about node degrees at each step to achieve such results. 
Herrera and Zufiria~\cite{2011Herrera} improved this process by using the number of steps in the random walks to guide triangle generation, introducing a way to control the network's clustering coefficient using again only local information.

%Herrera and Zufiria~cite{aaa} fizeram não sei o que. 
%Herrera et al.~cite{aaa} fizeram não sei o que. 
%Não sei o que foi proposto pelo Herrera et al.~cite{aaa}. 
%Estrategia X foi usada para conseguir resultado A~cite{aaa}. 

%Random walks are a prominent way to generate power-law networks. 

The random walk process proposed by Saramäki and Kaski~\cite{Saram_ki_2004} begins from a typically small initial graph with $n_0$ nodes. At each iteration, a new node $v$ is added to the graph, linking $v$ to existing nodes that will be chosen using random walks. 
%\st{where $m$ is the first algorithm parameter.} \rubenF{Não precisa, fica claro pelo contexto.} 
These chosen nodes (called ``marked nodes'' from now on) are identified as follows: beginning from a randomly selected node $w$, $l$ random steps are taken from $w$, allowing to revisit previous nodes. After each walk, the endpoint is marked, and the process continues until $m$ nodes are marked. The new node $v$ is then connected to marked nodes. The process finalizes after adding $N$ new nodes to the graph. 

Herrera and Zufiria~\cite{2011Herrera} noted that by changing the value of the parameter $l$, the number of steps in the random walk, it is possible to control the network's clustering coefficient. If $l = 1$, the neighbor of a marked node will also be marked, generating a triangle between these two neighbors and the added node, thus affecting the clustering coefficient. Each node $v$ has an associated value $p_v$, the probability of $l = 1$ if the random walk starts from that node. %, according to some probability distribution. 

Random walks controlling the walk length $l$ proved an efficient way to create power-law networks while regulating the clustering coefficient. However, the lack of attention to the average distances on the network remained an open question for real-world network generation using random walks. 

\section{Methods}

The primary goal of our algorithm is to generate long-tailed networks that combine high clustering coefficients with short path lengths between the nodes. 

Our algorithm starts from a small initial graph $G(V,E)$ with $n_0$ nodes. A new node $v$ is added to the graph at each iteration. Following Saramäki and Kaski~\cite{2011Herrera}, we perform a random walk, starting from a random node. We mark this initial node and then continue to mark each node reached after $l$ steps along the random walk, resulting in $m$ marked nodes. Edges from $v$ to each of the $m$ marked nodes are added to the graph at the end of the walk. 

To control the clustering coefficient, it can be used $l$, the number of steps between any two nodes marked successively. 
The value of $l$ is decided after marking some node, and before starting a new phase of the random walk, having some probability $p_1$ to be $1$, and probability $1-p_1$ to be 2. 
%where $p_1$ is another parameter. 

Note that $l$ follows a Bernoulli-like distribution, as proposed in~\cite{2011Herrera}, but it is possible to modify this distribution to contemplate larger values for $l$. 
%: $3, 4, 5$, and so on. 
However, as we will see later, we can already achieve the desired behavior with this simple distribution of $l$. 

At this point, unlike in~\cite{Saram_ki_2004,2011Herrera}, an additional edge is created at each iteration. 
The idea behind this step is to reduce the overall distances within the network (a neglected aspect in the original model) by adding shortcut edges, but not in an entirely random manner. 
The process begins by choosing a value $d$ with some probability $P(d)$. 
The distribution $P(d)$ is fixed for each algorithm execution, and decreases as $d$ grows in such a way that larger values of $d$ are less likely to be chosen than smaller ones. 
A random node $s$ in the network is then chosen, and we pick another node $t$ located at $d$ steps from $s$. % breadth-first search (BFS) é só uma forma de fazer
The new edge then connects $s$ and $t$. 
%If there is no node with distance d from s in the network, no edge is created in this iteration. 

The probability distribution we used is based on the idea that the likelihood of linking two nodes, $x$ and $y$, should decrease inversely proportional to the distance between $x$ and $y$ squared. 
In this way, shortcuts are created on the network, but not in a completely random manner, but somewhat closer to the link generation process in real networks, where more similar nodes are more likely to be connected. For example, two individuals who are closer to each other, in the geographical or topological sense, are more likely to meet. The emergence of these `random' edges reflects the natural appearance of new relations and links between nodes as the network grows. 

Thus, in our model $P(d) \propto \frac{1}{d^2}$, or $P(d) = \frac{A}{d^2}$ for some fixed $A$, since the probabilities for each $d$ value we use should sum to 1. The value of the normalizing constant $A$ may be found using the fact that 

$$A(\frac{1}{2^2} + \frac{1}{3^2} + ... + \frac{1}{d_{max}^2}) = 1,$$ 

where $d_{max}$ is the approximate current diameter of the graph. We estimate $d_{max}$, an approximation to the real diameter, in a simple way: assume that each node has a degree equal to the average degree $\overline{deg} = \frac{2|E|}{|V|}$ over all nodes. 
In an exponential branching process, at each distance $k$ from some node $v$, there are approximately $(\overline{deg}-1)^k$ nodes. Using the geometric series sum, there are at all $\frac{ (\overline{deg}-1)^k - 1}{\overline{deg}-2}$ nodes at a distance at most $k$ from $v$. 
% The nodes of the graph run out at some level $k$, being: 
Growing $k$, at some point, the number of covered nodes will reach the overall number of nodes $|V|$, being:  

$$ \frac{ (\overline{deg}-1)^k - 1}{\overline{deg}-2} = |V| $$

$$ (\overline{deg}-1)^k = |V| \cdot (\overline{deg}-2) + 1 $$

$$ k = \log_{\ \overline{deg}-1} \left( |V| \cdot (\overline{deg}-2) + 1 \right) $$

The sought diameter $d_{max}$ is then $ 2 \log_{\ \overline{deg}-1} \left( |V| \cdot (\overline{deg}-2) + 1 \right) $, twice the maximum value of $k$. 
%Since we used an upper bound of the real diameter of the network
The chosen value for $d$ is unlikely to be bigger than the real diameter of the network. 
%The probability distribution used here for choosing d is based on inverse squares:  

%The probability of a value $v$ being chosen for d is defined as: $P(d = v)$ = $P(A\sum_{i=2}^{v-1}1/{i^2} < a \leq A\sum_{i=2}^{v}1/{i^2}$), where $a$ is a random float between 0 and 1. 

In a simplified way, the proposed algorithm that builds the network goes as follows: 

\begin{enumerate}
    \item Start with a small initial graph with $n_0$ nodes.
    \item Add a new node $v$ to the network.
    \item Pick a random node $w$. Perform a random walk from $w$, marking each node reached after every $l$ steps. Stop when $m$ nodes are marked, and connect them to $v$. 
    % in random directions from $w$ and mark where you end up. Repeat this $m$ times, marking each ending node. 
    \item Choose a value $d$ with some probability $P(d)$ and a random node $s$. Find a node $t$ at a distance $d$ from $s$, and connect $s$ and $t$ with an edge. 
    \item Repeat $N$ times the steps 2 to 4. 
\end{enumerate}

%In short, the algorithm adds vertices to the graph, choosing their connections using random walks, and adding some controllable shortcut edges as the network grows. 

Algorithms~\ref{alg:aux} and~\ref{alg:main_alg} describe in more detail our implementation of the network generation approach presented in this paper. Algorithm~\ref{alg:aux} illustrates the random walk process that starts from a node $start$ in a graph $G$, uses the probability $p_1$ of $l$ being equal to $1$, and generates a list of $m$ marked nodes. 

In line 1, the RandomWalk algorithm initializes the marked list with the starting node, and in line 2, it sets the current node that tracks the position during the walk to $start$. Lines 3-8 are repeated $m - 1$ times, adding $m - 1$ nodes to the list of marked nodes. In line 4, the value of $l$ is established based on the parameter $p_1$, and in lines 5-7, the new phase of the random walk is performed. 
%in line 6, a random neighbor of the current node is chosen, and then this node becomes the current node. - Nível de detalhe alto demais. 
The endpoint is then added to the marked node list in line 8. 
Line 9 returns the list of marked nodes. 

\begin{algorithm}[H]
\caption{\textbf{RandomWalk} $(G, start, p_1, m)$}
\label{alg:aux}
\hspace*{\algorithmicindent}
\begin{algorithmic}[1]

\State $\text{marked} \gets \{\text{start}\}$  
\State $\text{current} \gets \text{start}$
\RepeatN{$m - 1$}
\State $\text{l} \gets \text{takes the value $1$ with probability $p_1$ and $2$ otherwise}$ 
          \RepeatN{$l$}
            \State $\text{neighbors} \gets \text{G[current].neighbors}$
            \State $\text{current} \gets \text{neighbors[randomIndex]}$
          \End
          \State $\text{marked} \gets \text{marked} \cup \{\text{current}\}$

    \End
    \State \Return marked

\end{algorithmic}
\end{algorithm}

On the other hand, Algorithm~\ref{alg:main_alg} describes the introduced network generation process. It takes as parameters an initial graph $G$, which is typically small, the number of nodes $N$ to be added to the graph, the probability $p_1$ of $l$ being equal to $1$, and the number of edges $m$ to be added at each iteration. 

The algorithm repeats $N$ times the following sequence of steps. It chooses a random node of the network as the starting point of the random walk in line 2. Then, the random walk is performed in line 3 by the procedure RandomWalk, returning the list of marked nodes. 
In line 4, a new node $v$ is added to the network, and in line 5, the edges are created between $v$ and the marked nodes. 
Lines 6-9 describe the process of adding an extra edge by choosing a distance $d$ based on some probabilistic distribution, and creating an edge between a random node $s$ and a node $t$ at $d$ steps from $s$. 
Line 10 returns the resulting network. 
 
\begin{algorithm}[H]
\caption{\textbf{GenerateNetwork} $(G, N, p_1, m)$}
\label{alg:main_alg}
\hspace*{\algorithmicindent}
\begin{algorithmic}[1]

  %\Comment{adjacency list of cycle graph with 10 nodes}
  %\State $\text{numberOfNodes} \gets \text{10}$ \red{Não precisa. O grafo pode ter simplesmente uma propriedade que é o seu número de nós.}
  
  \RepeatN{N}
      
      \State $\text{start} \gets \text{random node of the network}$

      \State $\text{marked} \gets \text{\textbf{RandomWalk}$(G, start, p_1, m)$}$

      \State $v \gets G.$\textbf{add\_node}$()$
      \For{$u \in \text{marked}$}
          $G.$\textbf{add\_edge}$(v, u)$
          
      \EndFor

      \State $\text{d} \gets \text{random value according to distribution $P(d)$}$

      \State $\text{s} \gets \text{random node of the network}$

      \State $\text{t} \gets \textbf{find\_node}(s, d)$ 

      \State $G.$\textbf{add\_edge}$(s, t)$
     
  \End
  \State \Return $G$
\end{algorithmic}
\end{algorithm}

% https://onlinelibrary.wiley.com/doi/full/10.1111/itor.12419 

%The time complexity of the algorithm is $O(N(N + m))$, since steps 10-26 are repeated N times, steps 16-17 and 19-20 both are repeated m times, and the BFS has time complexity proportional to the amount of nodes ($O(V+E)$, or $O(V)$ when the number of edges scales linearly with the number of nodes). This relatively low time complexity makes generating networks as large as 100.000 nodes possible in a feasible amount of time. 

\section{Results}

This section presents the main characteristics of the networks generated by the proposed algorithm. We show how it can be used to create graphs with characteristics similar to those found in real-world networks. In our experiments, we used as the initial graph $G$ a cycle with 10 nodes (circular graph, $C_{10}$). Each row in a table represents one execution of our algorithm. 

Our analysis focuses mainly on four key network measures: the average local clustering coefficient $\overline{C}$, which indicates the likelihood that two neighbors of a given node are also neighbors, averaged across all nodes in the network; global clustering coefficient $C$ (or transitivity), which is the ratio of closed triplets to the total number of triplets in the graph; the average shortest path length ($\overline{L}$), representing the mean shortest distance between any pair of nodes; the estimated power-law coefficient, denoted by $\gamma$, which characterizes the degree distribution of the network in the following way: the probability \(P(k)\) that a randomly selected node has degree \(k\) is approximately proportional to \(k^{-\gamma}\). The exponent is approximated by calculating the angular coefficient 
of the degree distribution on a log-log scale using the least squares method. 

%\rubenF{Colocar o grau máximo de cada rede?.. -- \blue{Só se rodasse todas as simulações novamente, não guardei esse dado durante as rodadas}.} 
%\rubenF{Expoente no intervalo entre 2 e 3 -- característica de redes reais.}

\begin{figure}[H]
    \centering
    \includegraphics[width=0.85\textwidth]{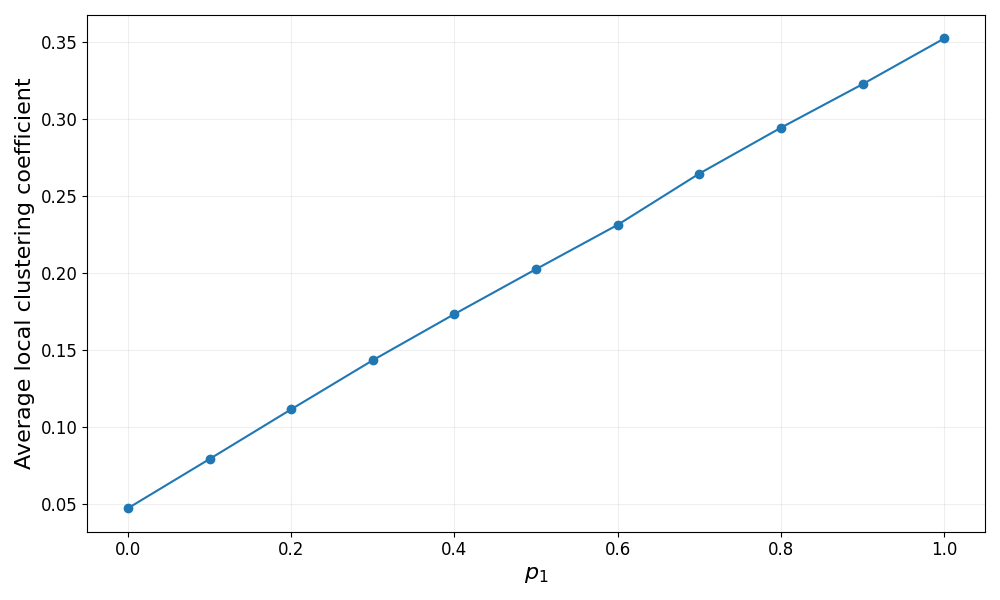}
    \caption{Controlling $\overline{C}$ by $p_1$: there is a linear relation between both measures ($m = 5$, $N = 50000$).}
    \label{fig:example_imageA}
\end{figure}

Our results show that the algorithm can generate a wide range of different networks. Table~\ref{tab:step_length_probabilities} shows that the clustering coefficients grow proportionally to the value of $p_1$, reaching a fairly high value for $p_1 = 1$, without affecting the other measures. This behavior can also be seen in Figure~\ref{fig:example_imageA}. Thus, by changing $p_1$ with a fixed $m$, it is possible to generate networks with different clustering coefficients, providing a simple way to control the value of this measure in the generated graphs. From Table~\ref{tab:step_length_probabilities}, we can also see that $\gamma$, the power-law exponent, varies little with $p_1$, staying almost constant after $p_1 = 0.3$. 

\begin{table}[H]
\centering
\caption{Generated networks and their measures using different $p_1$ values ($m = 5$, $N = 50000$). Parameters: probability $p_1$ of $l$ being equal to $1$. Measures: average local clustering coefficient $\overline{C}$; transitivity $C$; average shortest path length $\overline{L}$; estimated power-law exponent $\gamma$; maximum degree of the network $d_{max}$.}
\begin{tabular}{r@{\quad}r@{\quad}r@{\quad}r@{\quad}r@{\quad}r}
\hline
\multicolumn{1}{c}{\rule{0pt}{12pt} $p_1$} & 
\multicolumn{1}{c}{$\overline{C}$} & 
\multicolumn{1}{c}{$C$} & 
\multicolumn{1}{c}{$\overline{L}$} & 
\multicolumn{1}{c}{$\gamma$} & 
\multicolumn{1}{c}{$d_{max}$} \\[2pt]  \hline\rule{0pt}{12pt}

0.0 & 0.0461 & 0.0193 & 4.2705 & -1.9766 & 525 \\ 
0.1 & 0.0848 & 0.0310 & 4.2813 & -2.2508 & 382 \\ 
0.2 & 0.1207 & 0.0412 & 4.2889 & -1.9858 & 426 \\ 
0.3 & 0.1512 & 0.0501 & 4.2918 & -2.2084 & 448 \\ 
0.4 & 0.1814 & 0.0594 & 4.3115 & -2.2346 & 449 \\ 
0.5 & 0.2104 & 0.0680 & 4.3351 & -2.2049 & 417 \\ 
0.6 & 0.2390 & 0.0760 & 4.3494 & -2.1996 & 396 \\ 
0.7 & 0.2680 & 0.0841 & 4.3731 & -2.2039 & 546 \\ 
0.8 & 0.2971 & 0.0935 & 4.4125 & -2.1980 & 427 \\ 
0.9 & 0.3259 & 0.1033 & 4.4498 & -2.2459 & 403 \\ 
1.0 & 0.3549 & 0.1108 & 4.4762 & -2.2615 & 440 \\[2pt] \hline

\end{tabular}
\label{tab:step_length_probabilities}
\end{table}

\begin{figure}[H]
    \centering
    \includegraphics[width=0.85\textwidth]{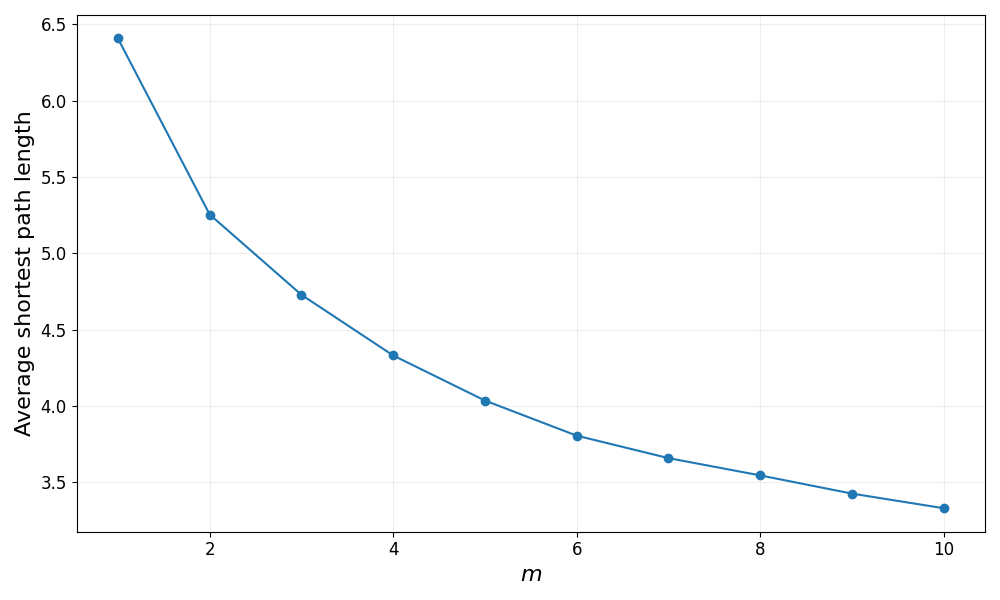}
    \caption{Controlling $\overline{L}$ by $m$: the average distance quickly drops down as $m$ increases ($p_1 = 0.5$, $N = 20000$).}
    \label{fig:example_imageB}
\end{figure}

Table~\ref{tab:marked_nodes_metrics} shows that both transitivity and the average shortest path length decrease as $m$ increases, while $\gamma$ remains relatively stable, fluctuating between -2.00 and -2.30 when $m \geq 2$.  Figure~\ref{fig:example_imageB} demonstrates this quick decay for the average shortest path length. 

\begin{table}[H]
\centering
\caption{Generated networks and their measures for different values of $m$ ($p_1 = 0.5$, $N = 20000$). Parameters: number of edges added at each iteration $m$. Measures: average local clustering coefficient $\overline{C}$; transitivity $C$; average shortest path length $\overline{L}$; estimated power-law exponent $\gamma$; maximum degree of the network $d_{max}$.}
\begin{tabular}{r@{\quad}r@{\quad}r@{\quad}r@{\quad}r@{\quad}r}
\hline
\multicolumn{1}{c}{\rule{0pt}{12pt} $m$} & 
\multicolumn{1}{c}{$\overline{C}$} & 
\multicolumn{1}{c}{$C$} & 
\multicolumn{1}{c}{$\overline{L}$} & 
\multicolumn{1}{c}{$\gamma$} & 
\multicolumn{1}{c}{$d_{max}$} \\[2pt]  \hline\rule{0pt}{12pt}

1  & 0.1358 & 0.0795 & 6.5044 & -2.5546 & 121 \\ 
2  & 0.3628 & 0.0951 & 5.3560 & -2.3339 & 202 \\ 
3  & 0.3239 & 0.0951 & 4.7796 & -2.0969 & 232 \\ 
4  & 0.2622 & 0.0847 & 4.3629 & -2.2439 & 282 \\ 
5  & 0.2125 & 0.0718 & 4.0473 & -2.0709 & 310 \\ 
6  & 0.1717 & 0.0635 & 3.8421 & -2.1094 & 348 \\ 
7  & 0.1467 & 0.0559 & 3.6765 & -2.1879 & 408 \\ 
8  & 0.1265 & 0.0508 & 3.5599 & -2.0339 & 444 \\ 
9  & 0.1123 & 0.0459 & 3.4487 & -2.0735 & 494 \\ 
10 & 0.1007 & 0.0416 & 3.3456 & -2.0741 & 505 \\[2pt] \hline

\end{tabular}
\label{tab:marked_nodes_metrics}
\end{table}

Finally, Table~\ref{tab:N_metrics} shows the characteristics of the generated networks for different values of $N$, the total number of added nodes, fixing $m = 5$ and $p_1 = 0.5$. 
It can be seen that after $N = 10.000$, the clustering coefficients vary little, remaining at a significantly high level. Table~\ref{tab:N_metrics} also shows the relationship between $\gamma$, and N, where the first decreases as the second increases. 

\begin{figure}[H]
    \centering
    \includegraphics[width=0.85\textwidth]{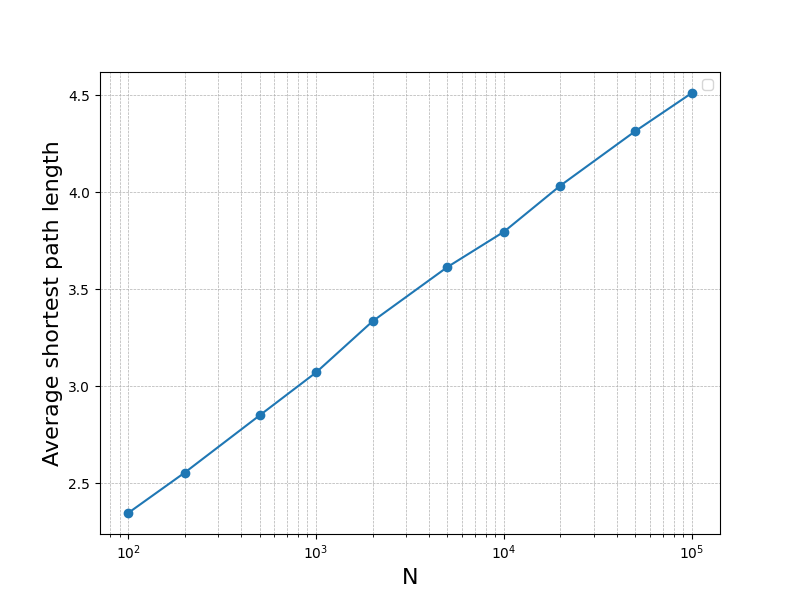}
    \caption{Relationship between the number of nodes $N$, plotted in logarithmic scale, and the average shortest path length $\overline{L}$ ($m = 5$, $p_1 = 0.5$).}
    \label{fig:example_imageC}
\end{figure}

Moreover, Figure~\ref{fig:example_imageC} and Table~\ref{tab:N_metrics} illustrate that the average shortest path length $\overline{L}$ grows proportionally to the logarithm of $N$ (small-world behavior), proving that the algorithm can create networks with short distances between the nodes. 

%\rubenF{Reforçar: temos o expoente no intervalo entre 2 e 3 -- característica das redes reais.}  

\begin{table}[H]
\centering
\caption{Generated networks and their measures for different values of $N$ ($m = 5$, $p_1 = 0.5$). Parameter: total number of nodes $N$ added to the graph. Measures: average local clustering coefficient $\overline{C}$; transitivity $C$; average shortest path length $\overline{L}$; estimated power-law exponent $\gamma$; maximum degree of the network $d_{max}$.}
\begin{tabular}{r@{\quad}r@{\quad}r@{\quad}r@{\quad}r@{\quad}r}
\hline
\multicolumn{1}{c}{\rule{0pt}{12pt} $N$} & 
\multicolumn{1}{c}{$\overline{C}$} & 
\multicolumn{1}{c}{$C$} & 
\multicolumn{1}{c}{$\overline{L}$} & 
\multicolumn{1}{c}{$\gamma$} & 
\multicolumn{1}{c}{$d_{max}$} \\[2pt]  \hline\rule{0pt}{12pt}

100     & 0.2949 & 0.2116 & 2.3636 & -0.9623 & 35 \\ 
200     & 0.2725 & 0.1668 & 2.5443 & -1.0640 & 48 \\ 
1000    & 0.2327 & 0.1144 & 3.1492 & -1.8068 & 89 \\ 
2000    & 0.2226 & 0.0974 & 3.3488 & -1.8760 & 136 \\ 
5000    & 0.2100 & 0.0675 & 4.3622 & -2.2204 & 433 \\ 
10000   & 0.2515 & 0.0807 & 4.4231 & -2.2239 & 594 \\ 
20000   & 0.2577 & 0.0852 & 4.4957 & -2.2362 & 484 \\ 
50000   & 0.2460 & 0.0842 & 4.6178 & -2.3176 & 559 \\ 
100000  & 0.2340 & 0.0796 & 4.7349 & -2.3260 & 678 \\[2pt] \hline

\end{tabular}
\label{tab:N_metrics}
\end{table}

For comparison, Table~\ref{tab:comparison_of_network_measures} shows the average local clustering coefficients and the average distances for networks generated by our algorithm and those generated without adding the special edge, equivalently to Herrera and Zufiria's model in which the nodes have a probability $p_1$ of making $l = 1$. The distances in our model are significantly lower, especially for smaller values of $m$, following the well-known \emph{six degrees of separation} principle. 
%, where the distance generated by the Herrera and Zufiria model is nearly twice as large. 

\begin{table}[H]
\centering
\caption{Comparison of networks generated by the Herrera and Zufiria model (1) and our model (2). Measures: average local clustering coefficient $\overline{C}$; average shortest path length $\overline{L}$.}
\begin{tabular}{r@{\quad}r@{\quad}r@{\quad}r@{\quad}r@{\quad}r}
\hline
\multicolumn{1}{c}{\rule{0pt}{12pt} $N$} & 
\multicolumn{1}{c}{$m$} & 
\multicolumn{1}{c}{$\overline{C_1}$} & 
\multicolumn{1}{c}{$\overline{C_2}$} & 
%\multicolumn{1}{c}{$C_1$} & 
%\multicolumn{1}{c}{$C_2$} & 
\multicolumn{1}{c}{$\overline{L_1}$} & 
\multicolumn{1}{c}{$\overline{L_2}$} \\[2pt]  
\hline\rule{0pt}{12pt}

%2000  & 5 & 0.2516 & 0.2162 & 0.1287 & 0.0978 & 3.8763 & 3.3346 \\

%5000  & 5 & 0.2457 & 0.2027 & 0.1053 & 0.0805 & 4.0319 & 3.6137 \\

%10000 & 5 & 0.2423 & 0.2057 & 0.1011 & 0.0727 & 4.2926 & 3.7956 \\

20000 & 2 & 0.4248 & 0.3734 & 10.2354 & 5.2538 \\
% & 0.1708 & 0.0944 

20000 & 3 & 0.4020 & 0.3286 & 7.3107 & 4.7691 \\

20000 & 4 & 0.3126 & 0.2505 & 6.2514 & 4.3309 \\
% & 0.1284 & 0.0813 

%20000 & 6 & 0.1913 & 0.1712 & 4.1499 & 3.8060 \\
% & 0.0805 & 0.0608

50000 & 2 & 0.4218 & 0.3694 & 11.4507 & 5.6787 \\
% & 0.1709 & 0.0889

50000 & 3 & 0.3960 & 0.3271 & 9.1531 & 5.0849 \\
% 0.1561 & 0.0892 &

50000 & 4 & 0.3130 & 0.2643 & 6.1754 & 4.6525 \\
% & 0.1227 & 0.0793 

70000 & 2 & 0.4200 & 0.3676 & 11.9726 & 5.8315 \\
% & 0.1705 & 0.0881 

70000 & 3 & 0.3963 & 0.3254 & 8.5344 & 5.1880 \\
% & 0.1562 & 0.0863 

70000 & 4 & 0.3140 & 0.2621 & 6.0804 & 4.7553 \\[2pt]
\hline
% 0.1213 & 0.0774 &

\end{tabular}
\label{tab:comparison_of_network_measures}
\end{table}

%\rubenF{Tempos de execução? \ldots Se conseguimos rebaixar o tempo quadrático, poderiamos pensar em colocar.} 

\section{Conclusions}

The study of complex networks gained prominence across various scientific disciplines~\cite{2023Boccaletti,2023Interianb,2021Scabini}. 
In this study, we presented a simple and flexible algorithm that generates a wide range of networks with different values of several network measures, such as clustering coefficients, average shortest path length, and the estimated power-law coefficient, without employing global information about the graph topology or degree distributions. 

The proposed approach can generate networks with long-tailed degree distributions, high clustering coefficients, and short average distances between the nodes, three of the fundamental characteristics of real-world networks. 
Using this simple random-walk-based approach, it is possible to generate networks with structural characteristics similar to those found in real-world networks without using hyperbolic geometry methods~\cite{2010Krioukov}. 
The implementation of our algorithm is publicly available on a GitHub repository~\cite{Morais2024repo}. 

We can control the average distances between network nodes, keeping them small and logarithmic in the size of the network, a neglected aspect in previous models~\cite{Saram_ki_2004,2011Herrera}. 
In addition, our model allows different probability distributions to set up the distance value $d$ used to introduce a degree of randomness in the network connections. By choosing the appropriate distribution, the average shortest path length $\overline{L}$ can be made smaller or larger, even for large graphs with tens and hundreds of thousands of nodes. 

Network clustering coefficients grow proportionally to the parameter $p_1$, reaching fairly high values while keeping realistic the other measures: average distances and ``long-tailness''. 

In future works, we intend to explore different distributions to control the value of $l$, the number of steps between any two nodes marked successively. We will study the possibility of increasing the values of $l$, exploring the impact of the chosen distribution on the clustering coefficients and the average distances between nodes. 

We are also interested in incorporating a fourth characteristic of real-world networks into our model: a high degree of modularity. High modularity implies the presence of modules (groups, communities) of nodes, with more dense connections within modules but sparse connections between nodes in different modules. Incorporating this feature organically, without being forced or planned in advance, seems like an interesting challenge to tackle. 

\section*{Acknowledgments}

Ruben Interian was supported by research grant PIND 2423/24 (Universidade Estadual de Campinas).

%
% ---- Bibliography ----
%

\bibliographystyle{unsrt}
\bibliography{main}

\end{document}